\RequirePackage{silence}
\documentclass[%
reprint,
superscriptaddress,
nofootinbib,
 amsmath,amssymb,
 aps,
pra,
floatfix,
]{revtex4-2}

\usepackage[T1]{fontenc}    
\usepackage{lmodern}
\usepackage[english,shorthands=off]{babel} 
\usepackage[table,usenames,dvipsnames]{xcolor}
\usepackage[colorlinks=true,
			hyperindex, 
		    linkcolor = NavyBlue,
		    anchorcolor = hyperrefblue,
		    citecolor = Green,
		    filecolor = hyperrefblue,
		    urlcolor = NavyBlue,
			breaklinks=true]{hyperref}
\usepackage{amsmath,amssymb}
\usepackage{amsthm}
\usepackage{quantikz}
\usepackage[capitalize]{cleveref} 
\crefname{figure}{Figure}{Figures} 
 
\usepackage{paralist}
\usepackage{enumitem}

\usepackage{url}            
\usepackage{booktabs}       

\usepackage{amsmath,amssymb,mathtools,dsfont} 
\usepackage{pifont}
\usepackage{etoolbox}

\usepackage{complexity}
\let\PP\undefined 
\let\EE\undefined

\usepackage{microtype}      
\usepackage{multirow}       
\usepackage{algorithm}
\usepackage{algpseudocode}
\algrenewcommand\algorithmiccomment[1]{\hfill $\triangleright$ #1} 
\usepackage{chemformula}
\let\ce\ch
\usepackage{graphicx}
\usepackage{grffile}
\usepackage{xcolor}
\usepackage{tikz}
\usepackage{pgfplots}
\DeclareUnicodeCharacter{2212}{−}
\usepgfplotslibrary{groupplots,dateplot}
\usetikzlibrary{positioning}
\usetikzlibrary{patterns,shapes.arrows}
\pgfplotsset{compat=newest}
\usepackage{tikzscale}
\usepackage{rotating}
\usepackage{nicefrac}

\usepackage[
nolist]{acronym}
\makeatletter
\AtBeginDocument{%
  \renewcommand*{\AC@hyperlink}[2]{%
    \begingroup
      \hypersetup{hidelinks}%
      \hyperlink{#1}{#2}%
    \endgroup
  }%
}
\makeatother

\renewcommand{\i}{\ensuremath\mathrm{i}}

\DeclareMathOperator{\LandauO}{\mathrm{O}}

\DeclareMathOperator{\Tr}{Tr}

\newcommand{\fro}{\mathrm{F}}

\DeclareMathOperator{\sign}{sign}

\DeclareMathOperator{\Var}{Var}

\NewDocumentCommand\Cl{mg}{
    \ensuremath{\mathrm{Cl}_{#1}\IfNoValueTF{#2}{}{(#2)}}%
}
\NewDocumentCommand\HW{mg}{
    \ensuremath{\mathrm{HW}_{#1}\IfNoValueTF{#2}{}{(#2)}}%
}

\newcommand{\RR}{\mathbb{R}}

\newcommand{\1}{\mathds{1}}
\newcommand{\EE}{\mathbb{E}}
\newcommand{\PP}{\mathbb{P}}

\newcommand{\mc}[1]{\mathcal{#1}}

\renewcommand{\vec}[1]{\boldsymbol{#1}}

\DeclarePairedDelimiterX{\abs}[1]{\lvert}{\rvert}{%
  \ifblank{#1}{\,\cdot\,}{#1}
}   

\DeclarePairedDelimiterX\norm[1]\lVert\rVert{%
  \ifblank{#1}{\,\cdot\,}{#1}
}   

\DeclarePairedDelimiterX{\iiiNorm}[1]{\lvert}{\rvert}{%
  \delimsize\lvert\delimsize\lvert#1\delimsize\rvert\delimsize\rvert%
}

\DeclarePairedDelimiterXPP\snorm[1]{}\lVert\rVert{_\infty}{\ifblank{#1}{\,\cdot\,}{#1}}   

\DeclarePairedDelimiterXPP\twonorm[1]{}\lVert\rVert{_2}{\ifblank{#1}{\,\cdot\,}{#1}}   

\DeclarePairedDelimiterXPP\trnorm[1]{}\lVert\rVert{_1}{\ifblank{#1}{\,\cdot\,}{#1}}   

\DeclarePairedDelimiterXPP\fnorm[1]{}\lVert\rVert{_{\fro}}{\ifblank{#1}{\,\cdot\,}{#1}}   

\DeclarePairedDelimiterXPP\dnorm[1]{}\lVert\rVert{_\diamond}{\ifblank{#1}{\,\cdot\,}{#1}}   

\DeclarePairedDelimiterXPP\cbnorm[1]{}\lVert\rVert{_\mathrm{cb}}{\ifblank{#1}{\,\cdot\,}{#1}}   
\DeclarePairedDelimiterXPP\onenorm[1]{}\lVert\rVert{_{1\rightarrow 1}}{\ifblank{#1}{\,\cdot\,}{#1}}   
\DeclarePairedDelimiterXPP\ddnorm[1]{}\lVert\rVert{_{\diamond\rightarrow \diamond}}{\ifblank{#1}{\,\cdot\,}{#1}}   
\DeclarePairedDelimiterXPP\ssnorm[1]{}\lVert\rVert{_{\infty\rightarrow\infty}}{\ifblank{#1}{\,\cdot\,}{#1}}   

\DeclarePairedDelimiterX\Set[1]\{\}{%
  
  #1
}

\DeclarePairedDelimiterX\innerp[2]{\langle}{\rangle}{%
  \ifblank{#1}{\,\cdot\,}{#1} , \ifblank{#2}{\,\cdot\,}{#2}%
}

\DeclarePairedDelimiterX\average[1]{\langle}{\rangle}{%
  \ifblank{#1}{\,\cdot\,}{#1}%
}

\DeclarePairedDelimiterX\ketbra[2]{\vert}{\vert}%
  {#1\kern0.15ex\delimsize\rangle\delimsize\langle\kern0.15ex\mathopen{}#2}

\DeclarePairedDelimiterX\sandwich[3]{\langle}{\rangle}%
  {#1\,\delimsize\vert\kern0.15ex\mathopen{}#2\kern0.15ex\delimsize\vert\kern0.15ex\mathopen{}#3}

\DeclarePairedDelimiterX\obraket[2]{(}{)}%
  {#1\kern0.15ex\delimsize\vert\kern0.15ex\mathopen{}#2}

\DeclarePairedDelimiterX\oketbra[2]{\vert}{\vert}%
  {#1\kern0.15ex\delimsize)\delimsize(\kern0.15ex\mathopen{}#2}

\DeclarePairedDelimiterX\osandwich[3]{(}{)}%
  {#1\,\delimsize\vert\kern0.15ex\mathopen{}#2\kern0.15ex\delimsize\vert\kern0.15ex\mathopen{}#3}

\newcommand{\hhu}{Faculty of Mathematics and Natural Sciences, Heinrich Heine University D{\"u}sseldorf, D{\"u}sseldorf, Germany}

\newcommand{\tuhh}{Institute for Quantum Inspired and Quantum Optimization, Hamburg University of Technology, Hamburg, Germany}

\newcommand{\eye}{\mathds{1}}
\newcommand{\real}{\mathbb{R}}

\begin{document}

\title{Reducing the sampling complexity of energy estimation in quantum many-body systems using empirical variance information}
\author{Alexander Gresch}%
\affiliation{\hhu}%
\affiliation{\tuhh}%
\author{Uğur Tepe}%
\affiliation{\hhu}%
\author{Martin Kliesch}%
\affiliation{\tuhh}%
\email{martin.kliesch@tuhh.de}%

\begin{abstract}
We consider the problem of estimating the energy of a quantum state preparation for a given Hamiltonian in Pauli decomposition. For various quantum algorithms, in particular in the context of quantum chemistry, it is crucial to have energy estimates with error bounds, as captured by guarantees on the problem's sampling complexity. 
In particular, when limited to Pauli basis measurements, the smallest sampling complexity guarantee comes from a simple single-shot estimator via a straightforward argument based on Hoeffding's inequality. 

In this work, we construct an adaptive estimator using the state's actual variance. 
Technically, our estimation method is based on the \ac{EBS} algorithm and grouping schemes, and we provide a rigorous tail bound, which leverages the state's empirical variance. 
In a numerical benchmark of estimating ground-state energies of several Hamiltonians, we demonstrate that \ac{EBS} consistently improves upon elementary readout guarantees up to one order of magnitude. 
\end{abstract}

\maketitle
\hypersetup{
pdftitle={Adaptive and provably accurate estimation of quantum expectation values using the empirical Bernstein stopping rule},
pdfsubject={Quantum many-body physics},
pdfauthor={Alexander Gresch, Ugur Tepe, Martin Kliesch},
pdfkeywords={Grouping,
				classical shadows,
				shadow, energy, estimation,
				randomized, measurements, measurement, grouping, 
				sample, complexity, 
				Pauli, strings, observables, 
				empirical, Bernstein, concentration, inequality, tail, bound, 
				quantum, many-body, Hamiltonian,  
				hybrid, variational, eigensolver, algorithm, VQE, VQA, 
				simulation, simulator, simulation, 
				readout, 
				adaptive, stopping, algorithm
				}
}
%

\section{Introduction}
\label{sec:introduction}

The most promising application area of quantum computers is arguably the simulation of physical systems, as initially envisioned by Feynman \cite{Fey82}. 
A particularly prominent example is the problem of calculating ground-state energies of quantum many-body Hamiltonians.
This is a basic, but practically important problem, e.g., in quantum chemistry and material science. 
However, classical computation methods suffer from the exponentially large dimension of the underlying Hilbert space.
To address this issue, the most competitive approaches rely on uncontrolled approximations. 
Hence, they are not supported by rigorous error bounds and yield unsatisfactory performance for strongly correlated systems. 
Quantum computing, on the other hand, promises to solve difficult quantum many-body problems much faster than would be possible otherwise~\cite{Hoefler23DisentanglingHypeFromB}. 
Moreover, several quantum algorithms have been proposed with rigorous guarantees and error bounds; see, e.g., Refs.~\cite {Liu2022ProspectsofQuantumComputing,Motta2022EmergingQuantumAlg} for an overview. 

Great efforts are being made to reduce the quantum hardware requirements to implement such quantum algorithms.
For example, reducing the quantum circuit depths as much as possible aims at mitigating the unavoidable hardware noise.
Another key step in this effort to lift the requirements is to read out the energy by simple measurements employing only short readout circuits rather than using, e.g., quantum phase estimation. 
This is particularly important for \ac{NISQ} algorithms, where much computational effort is made using classical computation, and only the classically most difficult parts are solved quantumly.
Nevertheless, the measurement subroutine now often constitutes a bottleneck in the quantum algorithm~\cite{Gonthier2022MeasurementRoadblock}: as a consequence of the direct readout, a large number of repeated measurements is required to accurately estimate the state's energy.
This roadblock is often tackled by efficient, yet often heuristic methods that are not accompanied by rigorous estimation guarantees, for example, in terms of sample complexities.
This is especially detrimental in situations where accuracy is key.
Furthermore, the optimal readout strategy is usually not known and devising suitable schemes has been subject of concentrated research over the last few years~\cite{Gokhale2019MinimizingStatePreparations,Jena2019PauliPartitioningWith,Verteletskyi2020MeasurementOptimizationVQE,Crawford2019EfficientQuantumMeasurement,Zhao2020MeasurementReductionVQE,arrasmith2020OperatorSampling}.

Current state-of-the-art methods often employ random measurements~\cite{Huang2019PredictingFeatures,Huang2020Predicting,Elben22RandomizedMeasurementToolbox}. 
To further reduce the measurement effort, several methods leverage the structure of the given Hamiltonian~\cite{Hadfield2020Measurements,Shlosberg2021AEQuO,Wu2021OverlappedGrouping,Gresch2023ShadowGrouping}.
These works still leave open the question of how the sampling complexity can be further reduced depending on the given state preparation. 
First attempts have been made to make use of the state's variance information \cite{Shlosberg2021AEQuO,Zhu2024OptimizingShotAssign}. Yet, in relevant settings, they cannot reach the accuracy of the state-agnostic ShadowGrouping method \cite{Gresch2023ShadowGrouping}. 
Nonetheless, in actual applications, we have to rely on rigorous quantities such as the sample complexity to gauge the efficacy of different readout methods.
As a consequence, we cannot use the state-of-the-art ShadowGrouping method and the currently best tail bound is provided by the single-shot estimator of Ref.~\cite{arrasmith2020OperatorSampling}.
The latter effectively corresponds to a worst case over all states as it completely discards their variance.

In this work, we propose to employ the \acf{EBS} algorithm~\cite{Mnih2008} to include the variance information gathered in a controlled fashion.
\Ac{EBS} incorporates the collected \emph{empirical} data adaptively to terminate the sampling process when the detected variance of the energy is low.
As a consequence, we benefit from an increased scaling with respect to the required precision and, hence, from a tighter tail bound when precision is paramount.
In a numerical benchmark on several qubit Hamiltonians inspired by the electronic structure problem, we find our algorithm consistently improving manifold over the currently tightest Hoeffding guarantee.

We structure the remainder of this manuscript as follows.
In \cref{sec:methods}, we give an overview of currently feasible state-of-the-art energy estimation procedures as well as available measurement guarantees associated.
This also includes an overview of the \ac{EBS} algorithm.
Afterwards, we explain how to employ it in the energy estimation context and follow up with a numerical benchmark in \cref{sec:results}.
We conclude with an outlook in \cref{sec:discussion}.
%
\section{Methodological Background}
\label{sec:methods}
In the following, we first revise how to measure a quantum state's energy on current quantum hardware with respect to a given Hamiltonian $H$ in \cref{subsec:estimation_task}.
The subsequent \cref{subsec:guarantees} introduces the tightest measurement guarantees to date.
Afterwards, we detail the \acf{EBS} algorithm in \cref{subsec:Bernstein_stopping_rule}, which capitalizes on a low empirical variance to reduce the total number of samples needed for an accurate estimation.
%
\subsection{The energy estimation task}
\label{subsec:estimation_task}
%
We wish to estimate the energy $E$ of a fermionic Hamiltonian, given by a quartic polynomial in the creation and annihilation operators of the considered orbitals, w.r.t.\ a fermionic quantum state, typically (an approximation to) its ground state or a lowly-excited state.
Since most quantum computing platforms require rephrasing the fermions into qubits, a first preparatory step consists of mapping the Hamiltonian to an equivalent qubit one using one of the many available fermion-to-qubit mappings to obtain an $n$-qubit Hamiltonian $H$.
The hardware then prepares the qubit state $\rho$.
In this setup, we are concerned with the determination of an estimate $\hat E$ of $E\coloneqq \Tr[\rho H]$.
To this end, we consider the Pauli decomposition of $H$ obtained from the fermion-to-qubit mapping, i.e.\
\begin{equation}
    \label{eq:Hamiltonian}
	H = \sum_{i=1}^M h_i O^{(i)}\, , \quad\quad O^{(i)} = \bigotimes_{j=1}^n O_j^{(i)}
\end{equation}
with $h_i \in \RR$ and single-qubit Pauli operators $O_j^{(i)} \in \{\1,X,Y,Z\}$.
The number of terms $M$ scales as $\LandauO(n^4)$ given a finite basis set of orbitals, which ensures the feasibility of the decomposition. 
Without loss of generality, we assume that no Pauli string $O^{(i)}$ is trivial, i.e., $O^{(i)} \neq \1^{\otimes n}\ \forall i$.

The probabilistic nature of quantum mechanics makes repeated measurements necessary for the energy estimation problem. 
Crucially, the measurement in the energy eigenbasis is infeasible in general.
Instead, one often measures the single terms $O^{(i)}$ to infer the energy estimate.
Since many Pauli terms commute with each other, it is beneficial to \emph{group} them into commuting families.
We refer to \cref{appendix:grouping} for further details on the associated grouping problem.
Subsequently, grouping a Hamiltonian produces $N_g$ disjoint index sets $\Set{\sigma_i}_{i=1}^{N_g}$ where $\sigma_i \subseteq [M]$ and $\bigcup_i \sigma_i = [M]$.
The disjointness constraint can be lifted but the corresponding theoretical treatment becomes much more involved~\cite{Gresch2023ShadowGrouping}.
Members of a group $\sigma$ can now be measured jointly as follows.
From a single measurement of $\rho$, we first determine each individual expectation value $o^{(i)}\coloneqq \Tr[\rho\, O^{(i)}]$ for each $i \in \sigma$.
Here, $\hat o^{(i)}$ denotes the outcome of a random variable $y_i\in\Set{-1,1}$ with respective outcome distribution given by Born's rule $\PP[y_i=1]=\Tr[\rho\,(O^{(i)}-\1)/2]$.
This procedure which we summarize in \cref{alg:grouped_estimator} is then repeated for each group returned by the grouping algorithm and yields a single-shot estimate for the state's energy $E$.
In total, this requires $N_g$ state preparations to yield a single sample for the state's energy.
The resulting \emph{grouped energy estimator} then reads as
\begin{equation}
	\hat{E} = \sum_{j=1}^{N_g} \sum_{i \in \sigma_j} h_i \hat{o}^{(i)} \,,
	\label{eq:energy_estimator}
\end{equation}
and recovers the state's energy in expectation, i.e.\ $\EE[\hat{E}] = E$.
Next, we wish to make sure that $\hat E$ is $\epsilon$-close to $E$ by repeating this routine sufficiently often.
\begin{figure}[t]
\centering
\input{plots/energy_estimator_algorithm}
\end{figure}
%
\subsection{Measurement guarantees}
\label{subsec:guarantees}
%
Given the energy estimator outlined in the previous section, we would like to supplement the output with
a rigorous appraisal of its accuracy.
That is, given the estimate $\hat E$, we want to quantify its closeness to the actual but unknown energy $E$ with high confidence.
For a given estimation error threshold $\epsilon > 0$, the \textit{failure probability}, i.e.\ that $\abs{\hat{E}-E} \geq \epsilon$ holds, captures this notion.
In general, this quantity cannot be efficiently evaluated (as it depends on the unknown quantum state produced in the experiment).
Nevertheless, we can often provide upper bounds to it that hold regardless of the quantum state under consideration.
Given a certain accuracy level, e.g.\ chemical accuracy, we aim for those bounds that require the least number of shots.
With \cref{eq:Hamiltonian}, Hoeffding's bound yields an upper bound to the total number of measurement rounds $N$ needed to reach accuracy $\epsilon$ with probability of at least $1- \delta$~\cite{arrasmith2020OperatorSampling}:
\begin{equation}
	N_\mathrm{Hoeff} \geq \frac{2}{\epsilon^2} \left( \sum_i \abs{h_i} \right)^2 \log\frac{2}{\delta}\,.
	\label{eq:single_shot_sample_complexity}
\end{equation}
We detail the corresponding, underlying measurement strategy in \cref{appendix:comparison_single_shot} as it differs from \cref{alg:grouped_estimator}.
While it offers the tightest rigorous tail bounds to date, this strategy, unfortunately, is not competitive with state-of-the-art grouping algorithms in practice.
In the latter, multiple samples are extracted from a single measurement outcome.
In general, this introduces correlated samples within \cref{eq:energy_estimator}, severely complicating an analytical treatment.
While this is practically possible, the 
corresponding guarantees still
cannot, to the best of our knowledge, compete with \cref{eq:single_shot_sample_complexity}~\cite{Gresch2023ShadowGrouping}.
Instead, we aim to leverage the \emph{empirical} variance of the energy estimator to improve on \cref{eq:single_shot_sample_complexity}.
This is the \acl{EBS} rule and explained in the following.
%
\subsection{The empirical Bernstein stopping algorithm}
\label{subsec:Bernstein_stopping_rule}
The guarantees from the previous section hold uniformly for all quantum states at once.
However, quantum states close to an eigenstate such as the ground state possess a low energy variance and, hence, need a decreased number of repeated measurement rounds to predict the energy to the same accuracy level.
Yet, the exact value of this variance, given a suitable measurement strategy determining the energy, is practically not available.
As a consequence, we have to rely on the \emph{empirical} variance information alone.
Intuitively, we seek to stop the measurement procedure prematurely if a low empirical variance is encountered repeatedly.
At the same time, we want to retain the same guarantee level in terms of the accuracy of the energy estimate.
The algorithm that combines both aspects is the \acf{EBS} algorithm~\cite{Mnih2008}.
It is an adaptive stopping algorithm that decides upon receiving samples of the state's energy whether to terminate the measurement procedure.
To this end, it keeps track of the empirical variance to profit from the \emph{empirical} Bernstein inequality~\cite{audibert2007}
\begin{equation} \label{eq:empirical_Bernstein_bound}
    \lvert \Bar E - E \rvert \leq \Bar{\sigma}_N\sqrt{\frac{2 \ln(3/\delta)}{N}} + R\frac{3 \ln(3/\delta)}{N} \eqqcolon \varepsilon_N \,,
\end{equation}
with empirical mean $\Bar{E}_N = \sum_{i=1}^N e_i / N$ 
and empirical variance $\Bar{\sigma}^2_N = \sum^N_{i=1}(e_i-\Bar{E}_N)^2/N$ after having collected $N$ energy samples $e_i$ obtained from $N$ independent and identical repetitions of \cref{alg:grouped_estimator}.
As a consequence, the range of \cref{eq:energy_estimator} is given by $R = 2 \sum_i \abs{h_i}$. 

If it terminates, \ac{EBS} is designed to yield an $\epsilon$-accurate result with a probability of at least $1 - \delta$.
Hence, the algorithm's inputs $(\epsilon,\delta)$ are referred to as the accuracy and the inconfidence, respectively.
Moreover, the total number of samples $\hat N$ required before terminating is ensured to be finite.
In fact, its expectation value can be bounded as
\begin{equation}
    \EE[\hat N] \leq C \max\left(\frac{\sigma^2}{\epsilon^2},\frac{R}{\epsilon}\right) \left[ \log \log \frac{R}{\epsilon} + \log \frac{3}{\delta} \right]\,,
\label{eq:EBS_sample_complexity}
\end{equation}
where $C$ is some constant independent of either $R$, $\epsilon$ or $\delta$ and where $\sigma^2 = \Var[\hat E]$ is the actual, unknown variance of the energy estimator.
For a very large variance, i.e.\ where $\sigma \lessapprox R$, \cref{eq:single_shot_sample_complexity,eq:EBS_sample_complexity} agree up to log-log factors.
However, in situations where $\sigma \ll R$, which is expected when estimating energies of states close to an eigenstate, \ac{EBS} promises up to a quadratic speed-up over the non-adaptive Hoeffding guarantee.
\begin{figure}[bt]
\centering
\input{plots/EBS_algorithm}
\end{figure}

To achieve this expected sample complexity~\eqref{eq:EBS_sample_complexity}, a few key concepts are incorporated into the algorithm.
At every step, the inconfidence $\delta$ in \cref{eq:empirical_Bernstein_bound} is replaced by a partial inconfidence $d_i > 0$.
The corresponding sequence $(d_i)_i$ is chosen such that its series converges to $\delta$, i.e.\ $\sum^\infty_{i=1} d_i = \delta$, to ensure that \ac{EBS} stops prematurely with probability at most $\delta$.
With this alteration, \ac{EBS} keeps gathering energy samples while tracking their running mean and empirical variance.
These values are then used to efficiently calculate $\varepsilon_N$ in \cref{eq:empirical_Bernstein_bound}.
If it drops below the accuracy parameter $\epsilon$, i.e.\ the \emph{stopping rule} is activated, \ac{EBS} terminates the sampling procedure.

Since the upper bound only decreases slowly, checking the stopping rule after every sample is wasteful.
Hence, as a second modification, \ac{EBS} employs geometric sampling, i.e.\ to iteratively increase the gap between subsequent stopping condition checks by a multiplicative factor $\beta > 1$.
If $(d_i)_i$ is furthermore chosen as a decreasing series, the smaller number of condition checks also increases the tolerated level of inconfidence at each check.
However, the geometric sampling may cause overshooting due to the exponentially growing gap between condition checks.
In order to mitigate this effect, an additional martingale-based, mid-interval stopping rule~\cite{mnih2008Master} is at hand that slightly modifies the tail bound~\eqref{eq:empirical_Bernstein_bound}.
We illustrate all aforementioned modifications as pseudocode in \cref{alg:EBS_Algorithm}.
\section{Results}
\label{sec:results}
%
We employ the \acf{EBS} algorithm of \cref{subsec:Bernstein_stopping_rule} to the energy estimation of the ground states of several quantum-chemistry-inspired Hamiltonians.
These Hamiltonians are derived from the paradigmatic electronic structure problem:
For a given molecule, we investigate how the electrons arrange themselves around the frozen nuclei at a given geometry.
This problem class has already sparked great interest of the quantum-computing community~\cite{Reiher2017,vonBurg2021} with tremendous experimental advances~\cite{Kandala17HardwareEfficientVariational,Hempel2018,Arute2020HartreeFock,Kawashima2021OptimizingElectronicStruct}.
In all of these approaches, a widespread subroutine of established workflows consists of the precise estimation of the energy of a quantum state from repeated measurements.
In this section, we first describe how to appropriately mend the \ac{EBS} algorithm in \cref{subsec:EBS_for_read_out}.
Afterwards, we evaluate it on qubit Hamiltonians inspired by the electronic structure problem of the \ce{H2}-molecule in \cref{subsec:numerics}.
To this end, we consider the dissociation curve, i.e., the ground-state energy $E(D)$ as a function of the interatomic distance $D$ and try to faithfully reconstruct it using repeated measurements.
In this numerical benchmark, we demonstrate that \ac{EBS} requires significantly fewer measurement rounds than mandated by the currently tightest readout guarantees~\cite{arrasmith2020OperatorSampling} based on Hoeffding's inequality~\eqref{eq:single_shot_sample_complexity}.
Finally, we demonstrate that this advantage pertains for larger Hamiltonians (in terms of their number of qubits) by investigation of the respective measurement rounds required by \ac{EBS} to accurately estimate the respective ground-state energies.
%
\subsection{Tailoring \ac{EBS} to the energy estimation}
\label{subsec:EBS_for_read_out}
%
We aim to apply the \ac{EBS} algorithm outlined in \cref{subsec:Bernstein_stopping_rule} to the energy estimation task using \cref{alg:grouped_estimator} as the subroutine to obtain iid. samples of the state's energy.
The algorithm is designed to stop when $\abs{\hat{E} - E} \leq \epsilon$ with probability at least $1 - \delta$.
However, there may be instances where the energy's variance is not small.
In this case, \ac{EBS} may actually inflict a measurement overhead which cannot be offset by a small empirical variance.
This motivates us to cap the maximal number of shots that \ac{EBS} may request.
Conveniently, \cref{eq:single_shot_sample_complexity} already provides an upper limit to the number of measurement rounds to achieve accuracy $\epsilon$ with probability at least $1 - \delta$.
As a consequence, we can make do with a finite sequence $(d_i)_i$ to partition the failure probability $\delta$ when checking the stopping condition.
Following \cref{alg:EBS_Algorithm}, we check at most $K$ times where $K$ fulfills
\begin{equation}
\begin{aligned}
    N_\mathrm{Hoeff} &\stackrel{!}{\geq} N_g \lfloor \beta^{K} \rfloor
    \\
    \Longrightarrow K &= \mc{O} \left( \log_\beta \frac{N_\mathrm{Hoeff}}{N_g} \right)
\end{aligned}
\end{equation}
Here, $N_g$ again denotes the number of groups and $\beta$ is a control parameter for the geometric sampling, see \cref{alg:EBS_Algorithm}.
Because we seek for highly accurate results, we also set the minimum number of samples \ac{EBS} requires conservatively to ten to reduce wasteful checks of \cref{eq:empirical_Bernstein_bound} and, hence, increase the remaining per-check confidences $d_i$ slightly.
Effectively, this means that we let $k_0 = \lceil \log_\beta 10 \rceil$.
Now, we distribute the total failure probability $\delta$ equally, that is $d_i = \delta / (K-k_0+1)$.
If \ac{EBS} does not terminate after $K$ checks, we can nevertheless stop sampling after $N_\mathrm{Hoeff}$ total measurement rounds to reach the guaranteed accuracy $\epsilon$.
We follow the energy estimation routine described in this section throughout all numerical studies.
\begin{figure}[tb]
    \centering
    \input{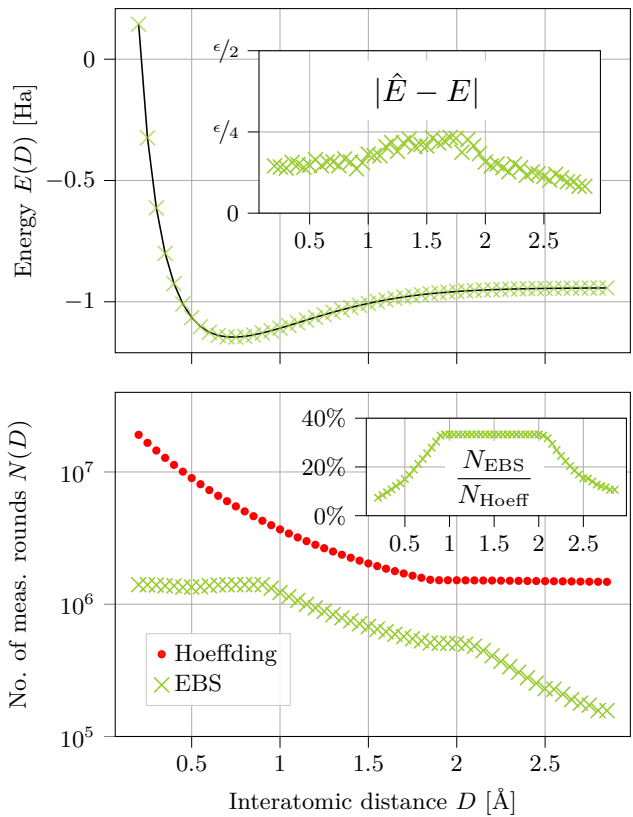}
    \caption{Efficiently estimating the dissociation curve of \ce{H2}.
    \textbf{Above}: for all considered interatomic distances $D$ (green), we reliably reach $E_\mathrm{GS}(D)$ (black line) up to chemical accuracy $\epsilon$.
    The inset shows the final error $\abs{\hat E - E}$ normalized by $\epsilon$. As the geometric sampling of \ac{EBS} can likely result in overshooting, the actual accuracy achieved is four times smaller than $\epsilon$.
    \textbf{Below}: the associated number of measurement rounds $N(D)$ for the readout of $E(D)$.
    \Ac{EBS} (green) improves upon the sample complexity~\eqref{eq:single_shot_sample_complexity} provided by Hoeffding's inequality (red dots) significantly for all choices of $D$.
    The inset shows that \ac{EBS} requires only $10-30\%$ of the latter.}
    \label{fig:H2_curve}
\end{figure}
%
%
\subsection{Numerical results}
\label{subsec:numerics}
%
%
\begin{figure}[b]
    \centering
    \input{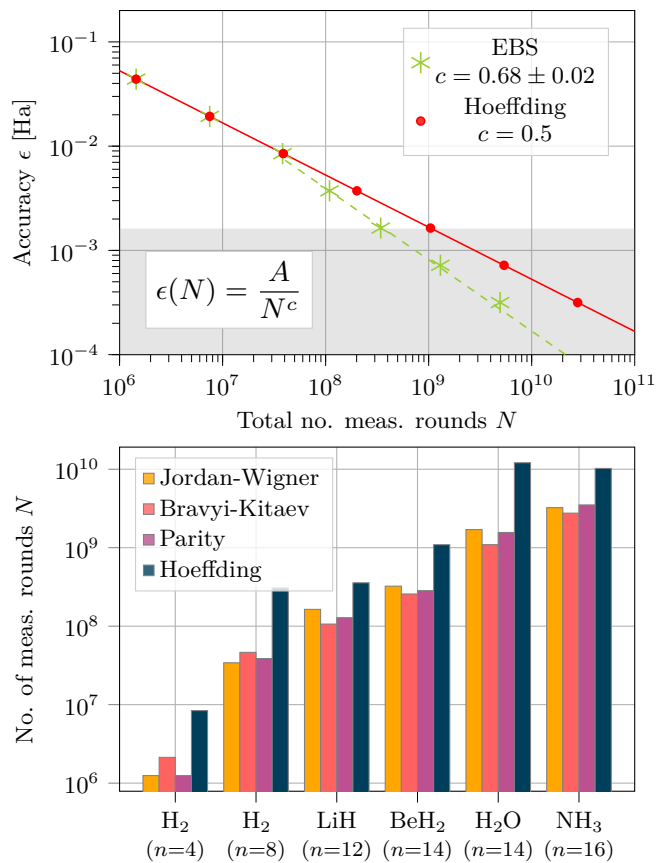}
    \caption{Estimation of ground-state energies for Hamiltonians inspired by various molecules using \ac{EBS}.
    \textbf{Above}: Reliable accuracy $\epsilon$ achieved by \ac{EBS} (green asterisks) and the single-shot estimator (red dots) for \ce{BeH2}-inspired Hamiltonians (in \ac{JW} encoding) as a function of the total number of measurement rounds $N$.
    The respective lines correspond to power-law fits to the data as per \cref{eq:fit_func}.
    Eventually, the favorable scaling of \ac{EBS}~\eqref{eq:EBS_sample_complexity} kicks in, leading to a decreased required measurement effort.
    \textbf{Below}: Sample complexities to estimate ground-state energies up to chemical accuracy for various molecule-inspired Hamiltonians and fermion-to-qubit mappings .
    Hoeffding's guarantee~\eqref{eq:single_shot_sample_complexity} is independent of the chosen mapping.} 
    \label{fig:GS_Molecules}
\end{figure}
We benchmark our adapted \ac{EBS} algorithm, \cref{alg:EBS_Algorithm} with modifications as per \cref{subsec:EBS_for_read_out}, on Hamiltonians inspired by the electronic structure problem of the \ce{H2} molecule; see \cref{appendix:vqe_details} for details on the choice of basis set, encoding and the readout procedure.
Repeating the estimation procedure for various interatomic distances $D$ yields the energy dissociation curve depicted in \cref{fig:H2_curve}.
Furthermore, we also magnify the actual error in multiples of $\epsilon$ as an inset.
As promised by \ac{EBS}, we obtain inaccurate results, i.e.\ with deviation larger than chemical accuracy, with probability at most $\delta = 10\%$.
In fact, we find that \ac{EBS} consistently yields $\abs{\hat E - E} \leq \epsilon/4$ empirically.
Additionally, we show each individual number of measurement rounds $N(D)$ required for the accurate readout.
We contrast this to the previous Hoeffding-based guarantee~\eqref{eq:single_shot_sample_complexity}.
In particular, for the smallest and largest considered interatomic distances, \ac{EBS} requires significantly fewer samples than commanded by Hoeffding's inequality~\eqref{eq:single_shot_sample_complexity}.
The kink in the $N(D)$-curves of \cref{fig:H2_curve}  is attributed to the range $R(D)$ of the respective energy estimators.
In the inset, we plot the percentage of samples $N_\mathrm{EBS}/N_\mathrm{Hoeff}$ \ac{EBS} requires to reliably reach the same promised accuracy.
This quantity does only weakly depend on the range (cf.~\cref{eq:EBS_sample_complexity}) and highlights the roughly fourfold shot reduction when compared against the Hoeffding guarantee.

Following up on this small \ce{H2}-example, we now show numerically that the advantage of \ac{EBS} over the Hoeffding bound persists also for systems represented by a larger qubit number $n$.
Due to the construction of the grouped energy estimator~\eqref{eq:energy_estimator}, this is a delicate question:
the energy estimator requires all groups $N_g$ to be measured at least once.
However, the number of terms in the decomposition $M$ grows as $\mc{O}(n^4)$ with $N_g$ possessing a more complicated dependence on $n$.
As a consequence, both sample complexities of \ac{EBS} and the single-shot estimator will also grow polynomially with $n$, obfuscating their exact scaling.
We therefore quantify the advantage of \ac{EBS} for various $n$ as follows.
We generate further fermionic Hamiltonians at around equilibrium bond lengths using Qiskit~\cite{Qiskit}.
Furthermore, we select three standard fermion-to-qubits mappings (\ac{JW}, \ac{BK} and the Parity transformation) that output the corresponding Pauli decomposition~\eqref{eq:Hamiltonian}.
In order to efficiently partition the decomposition into commuting groups, we employ ShadowGrouping~\cite{Gresch2023ShadowGrouping} until each observable in the decomposition is member of exactly one group.
This readily yields the measurement circuits and energy estimator~\eqref{eq:energy_estimator} necessary to determine the state's energy.
Afterwards, we exactly diagonalize each Hamiltonian to obtain the ground state.
Next, we use \ac{EBS} in conjunction with the estimator to measure the state's energy accurately up to chemical accuracy.
All sampling has been carried out using qibo~\cite{qibo}.
The code generating our findings can be found in a online repository~\cite{ebs_git_repo}.
\Cref{fig:GS_Molecules} summarizes our findings.
We exemplarily show the reached accuracy $\epsilon$ as a function of the total number of measurement rounds $N$ in estimating the ground-state energy of a quantum Hamiltonian inspired by the \ce{BeH2} molecule in \ac{JW}-encoding which is represented by $n=12$ qubits.
Again, the single-shot estimator obtains an energy estimate from each measurement outcome whereas \ac{EBS} aggregates $N_g = 159$ outcomes (one for each of the groups in \cref{eq:energy_estimator}) for a single estimate.
We choose seven different values for $\epsilon$ logarithmically equidistant between $10^{-4}$ and $10^{-1}$ and run \ac{EBS} with parameter $\delta=10\%$ for each $\epsilon$ and record $N(\epsilon)$.
Given the sample complexities, \cref{eq:EBS_sample_complexity,eq:single_shot_sample_complexity}, we know that $\epsilon(N)$ follows a power law of the form
\begin{equation}
    \epsilon(N) = \frac{A}{N^c}\,,
    \label{eq:fit_func}
\end{equation}
with some constant $A$ and, more importantly, coefficient $c$, determining the slope in \cref{fig:GS_Molecules}.
Hoeffding's inequality ensures that $c=1/2$ for the single-shot estimator.
By contrast, \ac{EBS} exhibits a value significantly larger for the smallest probed accuracy parameters $\epsilon \lesssim 10^{-2}$.
This can be attributed to the $\max(\sigma^2/\epsilon^2,R/\epsilon)$-term in \cref{eq:EBS_sample_complexity} because measuring in Pauli bases rather than the energy eigenbasis inflicts a variance $\sigma^2 > 0$.
As a result, the additional constant overhead $A$ of \ac{EBS} nullifies any advantage for moderate accuracy levels $\epsilon \gtrsim 10^{-2}$.
Since quantum chemistry applications require highly accurate estimates, however, \ac{EBS} gains an edge over the single-shot estimator.
Lastly, we run \ac{EBS} directly for chemical accuracy of $\epsilon \approx 1.6$ mHa and $\delta=10\%$ for various molecules and encodings.
In all considered cases, we observe that \ac{EBS} improves upon the single-shot estimator severalfold, even when considering Hamiltonians of increased system size $n$.
Our results give justified hope to the viability of employing \ac{EBS} in more sophisticated quantum chemistry applications where both efficiency and accuracy are paramount.
\section{Discussion and Outlook}
\label{sec:discussion}
In this work, we consider the energy estimation task often encountered in current and near-term quantum algorithms.
Importantly, we put emphasis on the associated readout guarantees.
To this end, we propose to employ the \acf{EBS} algorithm.
By continuously leveraging the \emph{empirical} data being collected, the algorithm decreases the overall measurement effort if a low empirical variance is encountered.
For our numerical benchmarks, inspired by crucial subroutines in quantum chemistry applications such as determining ground-state properties, we first provide a readout scheme that is compatible with \ac{EBS} and is based on grouped measurements.
In our numerical findings, this scheme consistently improves up to tenfold over the current best strategy in terms of the sample complexity sufficient to precisely estimate the ground-state energies of several Hamiltonians.
Importantly, this advantage appears to pertain even when considering Hamiltonians of increasing system size.
This is an encouraging step towards efficient readout algorithms that are fit to meet high-precision requirements, such as those often encountered in quantum chemistry applications. 
Furthermore, this includes, as one immediate use-case, the optimization and final readout of hybrid quantum algorithms such as the \ac{VQE}~\cite{PerMcCSha14,Wecker2015,McCRomBab16}.

Our proposed readout strategy may be further improved in future work.
For one, the energy sampling procedure of \cref{alg:grouped_estimator} is only one possibility to deal with a grouped Hamiltonian.
In general, both the grouping and the shot allocation to the several groups are not optimal.
Since \ac{EBS}' sample complexity~\eqref{eq:EBS_sample_complexity} directly depends on the variance of the energy estimator, selecting groups based on their individual (empirical) variance (or shot allocation in the spirit of Refs.~\cite{Shlosberg2021AEQuO,Zhu2024OptimizingShotAssign}) may help decrease the sample complexity further.
In this regard, we have explored to employ the single-shot estimator~\cite{arrasmith2020OperatorSampling} (\cref{alg:singleshot_estimator} in \cref{appendix:comparison_single_shot}) used to derive \cref{eq:single_shot_sample_complexity} in conjunction with \ac{EBS}.
However, we found that the additional variance introduced by sampling the Pauli term to measure next (almost) completely offsets any performance gains from employing \ac{EBS}.
Nevertheless, we leave refinements of a deterministic readout strategy to future works.
Another improvement can be achieved by employing deeper measurement circuits, depending on the quantum Hamiltonian under consideration~\cite{eckstein2024shotnoisereduction}.
Finally, we are currently trying to elevate the empirical Bernstein inequality~\eqref{eq:empirical_Bernstein_bound} to more complex structures such as random matrices.
This possibly enables us to mend \ac{EBS} to state-of-the-art readout methods that neither impose a fixed grouped Hamiltonian~\cite{Gresch2023ShadowGrouping} nor require disjoint groups~\cite{Wu2021OverlappedGrouping}.
\begin{acknowledgments}
This work has been funded by 
the Deutsche Forschungsgemeinschaft (DFG, German Research Foundation) via the Emmy Noether program (Grant No.\ 441423094); 
the Federal Ministry of Education and Research (BMBF) within the Research Program Quantum Systems via the joint projects QUBE (grant number 13N17149); and by 
the Fujitsu Germany GmbH and Dataport as part of the endowed professorship ``Quantum Inspired and Quantum Optimization''. 
\end{acknowledgments}
\subsection*{Author contributions}
A.G.\ carried out all relevant calculations and designed the numerical studies, U.T. implemented and carried out the numerical studies, M.K.\ supervised the process and conceived the idea of applying \ac{EBS} in the quantum context. 
All authors wrote the manuscript together. 
The authors declare no competing financial interest.
%
\subsection*{Data availability}
The Hamiltonians for the \ce{H2}-molecule are tabulated in Ref.~\cite{OMalley2015}.
The Hamiltonian decompositions used for \cref{fig:GS_Molecules} in \cref{subsec:numerics} have been sourced from an online repository~\cite{Hamiltonians_git_repo}.
The source code of ShadowGrouping to partition the respective Pauli decompositions has been taken and altered from Ref.~\cite{ShadowGrouping_git_repo}.
Any intermediate data generated for the benchmarks is stored alongside the computer code in Ref.~\cite{ebs_git_repo}.

\begin{acronym}[POVM]
\acro{ACES}{averaged circuit eigenvalue sampling}
\acro{AGF}{average gate fidelity}
\acro{AP}{Arbeitspaket}

\acro{BK}{Bravyi-Kitaev}
\acro{BOG}{binned outcome generation}

\acro{CNF}{conjunctive normal form}
\acro{CP}{completely positive}
\acro{CPT}{completely positive and trace preserving}
\acro{cs}{computer science}
\acro{CS}{compressed sensing} 
\acro{ctrl-VQE}{ctrl-VQE}

\acro{DAQC}{digital-analog quantum computing}
\acro{DD}{dynamical decoupling}
\acro{DFE}{direct fidelity estimation} 
\acro{DFT}{discrete Fourier transform}
\acro{DL}{deep learning}
\acro{DM}{dark matter}

\acro{EBS}{empirical Bernstein stopping}

\acro{FFT}{fast Fourier transform}

\acro{GS}{ground-state}
\acro{GST}{gate set tomography}
\acro{GTM}{gate-independent, time-stationary, Markovian}
\acro{GUE}{Gaussian unitary ensemble}

\acro{HOG}{heavy outcome generation}

\acro{irrep}{irreducible representation}

\acro{JW}{Jordan-Wigner}

\acro{LBCS}{locally-biased classical shadow}
\acro{LDPC}{low density partity check}
\acro{LP}{linear program}

\acro{MAGIC}{magnetic gradient induced coupling}
\acro{MAX-SAT}{maximum satisfiability}
\acro{MBL}{many-body localization}
\acro{MIP}{mixed integer program}
\acro{ML}{machine learning}
\acro{MLE}{maximum likelihood estimation}
\acro{MPO}{matrix product operator}
\acro{MPS}{matrix product state}
\acro{MS}{M{\o}lmer-S{\o}rensen}
\acro{MSE}{mean squared error}
\acro{MUBs}{mutually unbiased bases} 
\acro{mw}{micro wave}

\acro{NISQ}{noisy and intermediate scale quantum}

\acro{ONB}{orthonormal basis}
\acroplural{ONB}[ONBs]{orthonormal bases}

\acro{POVM}{positive operator valued measure}
\acro{PQC}{parametrized quantum circuit}
\acro{PSD}{positive-semidefinite}
\acro{PSR}{parameter shift rule}
\acro{PVM}{projector-valued measure}

\acro{QAOA}{quantum approximate optimization algorithm}
\acro{QC}{quantum computation}
\acro{QEC}{quantum error correction}
\acro{QFT}{quantum Fourier transform}
\acro{QM}{quantum mechanics}
\acro{QML}{quantum machine learning}
\acro{QMT}{measurement tomography}
\acro{QPT}{quantum process tomography}
\acro{QPU}{quantum processing unit}
\acro{QUBO}{quadratic binary optimization}
\acro{QWC}{qubit-wise commutativity}

\acro{RB}{randomized benchmarking}
\acro{RBM}{restricted Boltzmann machine}
\acro{RDM}{reduced density matrix}
\acro{rf}{radio frequency}
\acro{RIC}{restricted isometry constant}
\acro{RIP}{restricted isometry property}
\acro{RMSE}{root mean squared error}

\acro{SDP}{semidefinite program}
\acro{SFE}{shadow fidelity estimation}
\acro{SIC}{symmetric, informationally complete}
\acro{SPAM}{state preparation and measurement}
\acro{SPSA}{simultaneous perturbation stochastic approximation}

\acro{TT}{tensor train}
\acro{TM}{Turing machine}
\acro{TV}{total variation}

\acro{VQA}{variational quantum algorithm}
\acro{VQE}{variational quantum eigensolver}

\acro{XEB}{cross-entropy benchmarking}

\end{acronym}
\bibliographystyle{./myapsrev4-2}
\newcommand{\prr}{Phys.\ Rev.\ Research}
\bibliography{ag,mk}
\newpage
\newpage
\section*{Appendix}
\appendix
\renewcommand{\thesubsection}{\Alph{subsection}}
\setcounter{equation}{0}
\renewcommand{\theequation}{S\arabic{equation}} 
%
This appendix recaps relevant strategies for the direct readout of a quantum state's energy.
To this extent, we briefly revise grouping schemes that aim to achieve a measurement reduction in \cref{appendix:grouping}.
Afterwards, in \cref{appendix:comparison_single_shot}, we follow up with a rigorous definition of the single-shot estimator introduced in \cref{subsec:guarantees} and used in the numerical benchmark in \cref{subsec:numerics}.
We conclude with further details on how the Hamiltonians of the numerical benchmark have been derived in \cref{appendix:vqe_details}.
%
\subsection{Grouping methods}
\label{appendix:grouping}
%
In the Hamiltonian decomposition~\eqref{eq:Hamiltonian}, many Pauli terms commute with each other and can, hence, be measured simultaneously.
Such a set of pairwise commuting Pauli terms is called a group.
In order to ensure a resulting unbiased estimator, we have to find a partitioning of the terms into $N_g$ commuting groups.
This is referred to as the grouping of Pauli observables and reduces the number of distinct measurements from $M$ down to $N_g$.
Grouping schemes can take into account the coefficients of the terms~\cite{Jena2019PauliPartitioningWith,Gokhale2019MinimizingStatePreparations} or even the empirical data from previous measurements~\cite{Shlosberg2021AEQuO,Zhu2024OptimizingShotAssign}.
In general, the efforts strive towards minimizing $N_g$, which constitutes, in its decision version, an \NP-hard problem w.r.t.\ the system size~\cite{Wu2021OverlappedGrouping}.

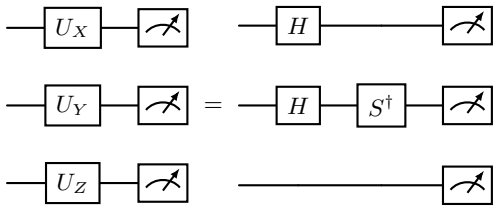
\begin{figure}[bt]
       \begin{quantikz}
           &\gate{U_X}& \meter{} \\
           &\gate{U_Y}& \meter{} \\
           &\gate{U_Z}& \meter{}
       \end{quantikz}
       =
       \begin{quantikz}
           \ghost{U_X} & \gate{H} & \ghost{S^\dagger}  & \meter{} \\
           \ghost{U_Y} & \gate{H}  & \gate{S^\dagger}  & \meter{} \\
           \ghost{U_Z} & \ghost{H} & \ghost{S^\dagger} & \meter{}
       \end{quantikz}
   \caption{%
   The three single-qubit Pauli basis measurement circuits $\{ U_X, U_Y, U_Z \}$ of depths at most two. $H$ and $S$ denote the Hadamard and the phase gate, respectively.
   }
   \label{fig:readout_circuits}
\end{figure}

Finally, grouping Pauli terms together requires additional measurement circuitry.
While these can be efficiently constructed from a given group~\cite{Gokhale2019MinimizingStatePreparations}, the resulting circuits require several two-qubit gates in general.
Constraining the circuits to consist only of single-qubit gates is also possible.
In this case, the measurement circuit consists of an appropriate Pauli basis measurement as depicted in \cref{fig:readout_circuits}.
However, it demotes the general commutativity to \ac{QWC}: two Pauli strings commute qubit-wise if they commute on all qubits simultaneously.
\Ac{QWC} implies general commutativity, but the contrary is not true, potentially curtailing the efficacy of the grouping approach.
Nevertheless, we opted for \ac{QWC} to group the various Hamiltonians due to its near-term applicability, for example, concerning noise overheads when implementing the readout circuits on actual hardware.
With this constraint, an estimate of $\Tr[\rho O]$ for the Pauli observable $O$ can be extracted from $N$ bit strings $\vec b^{(i)} \in \{\pm 1\}^n$ by
\begin{equation}
    \hat O \coloneqq \frac{1}{N} \sum_{i=1}^N \prod_{j: O_j \neq \1} b_j^{(i)}\,,
    \label{eq:Pauli_estimator}
\end{equation}
provided that each of the $N$ Pauli bases share \ac{QWC} with $O$.
%
\subsection{Details on the single-shot estimator}
\label{appendix:comparison_single_shot}
%
In \cref{sec:results}, we have relied on measuring groups of Pauli terms jointly to gather energy samples for the \ac{EBS} algorithm.
In the numerical benchmark, we compare this approach to an alternative one to which we refer as the single-shot estimator.
It comes along with its own readout guarantee~\eqref{eq:single_shot_sample_complexity}
of which we now give a short derivation.
To this end, we employ the Weighted Random Sampling method of Ref.~\cite{arrasmith2020OperatorSampling}.
In this method, we sample the next measurement setting based on the terms in the Pauli decomposition.
Crucially, the relative magnitude of the observable's coefficient determines its probability of being picked, i.e.\
$p_i = \abs{h_i}/\sum_j \abs{h_j}$ to measure only the $i$-th observable.
As a consequence, each measurement outcome serves as an estimate for the state's energy; see \cref{alg:singleshot_estimator} for its pseudocode.

Moreover, the corresponding estimator comes with rigorous guarantees.
Assume we have picked the $k$-th observable to be measured and received outcome $\hat{o}^{(k)}$.
The single-shot estimator $\hat{E}$ is then defined as
\begin{equation}
\begin{aligned}
	\hat{E} &= s_k \sum_{i=1}^M\abs{h_i} \\
	\text{where}\ s_k &\coloneqq \sign(h_k) \hat{o}^{(k)} \in \Set{\pm 1}\,.
\end{aligned}
\label{eq:single_shot_definition}
\end{equation}
It is unbiased and fulfills $\abs{\hat{E}} \leq \sum_i \abs{h_i}$.
Finally, Hoeffding's inequality tells us that with probability at least $1 - \delta$ (and $0 < \delta < 1/2$),
\begin{equation*}
	N \geq \frac{2}{\epsilon^2} \left( \sum_i \abs{h_i} \right)^2 \log\frac{2}{\delta}
\end{equation*}
samples (i.e.\ independent repetitions of \cref{alg:singleshot_estimator}) suffice for the empirical mean $\hat{E}_N = \sum_{i=1}^N \hat{E}^{(i)}/N$ to ensure $\abs{\hat{E}_N - E} \leq \epsilon$, verifying \cref{eq:single_shot_sample_complexity}.
%
\subsection{Details on the numerical benchmark}
\label{appendix:vqe_details}
%
We benchmark \ac{EBS} w.r.t.\ the energy estimation on the paradigmatic example of the electronic structure problem, i.e.\ the specific configuration of the electrons within a molecule.
For ease of presentation, we illustrate this workflow for obtaining a Hamiltonian inspired by the electronic structure problem of the \ce{H2}-molecule.
To this extent, we first obtain the fermionic Hamiltonian in the second quantization (including the Born-Oppenheimer approximation at fixed bond length $D$ between the two nuclei).
As the orbital basis set, we chose the minimal STO-3G, and we default to this choice throughout this work unless specifically indicated otherwise.
Afterwards, we map the fermionic Hamiltonian to a qubit one~\eqref{eq:Hamiltonian} using one of several possible fermion-to-qubit mappings such as the \ac{JW}~\cite{JordanWigner}, the \ac{BK}~\cite{BravyiKitaev} or the Parity transformation~\cite{BravyiKitaev,ParityTransformation}.
In the case of Parity encoding, we can remove redundant qubits further by considering symmetry. Afterwards, the resulting qubit Hamiltonian consists of two qubits and six Pauli terms:
\begin{equation} \label{VQE_Hamiltonian}
    H= \left( g_1 Z\eye +g_2 \eye Z +g_3 ZZ \right) + g_4 YY+ g_5 XX\,,
\end{equation}
where $g_\alpha(D) \in \real$ for each interatomic distance $D$ and we have omitted the offset energy $g_0(D)$.
These can be grouped into $N_g = 3$ disjoint groups (grouped terms are indicated by parenthesis), which require single-qubit Pauli basis measurements only.
The other Hamiltonians investigated in \cref{fig:GS_Molecules} can be obtained analogously.
There, we use both the STO-3G as well as the 6-31G basis sets for the \ce{H2}-Hamiltonians with four and eight qubits, respectively. 

We now seek to estimate its ground-state energy $E(D)$ using our \ac{EBS} subroutine to ensure accurate energy estimation up to chemical accuracy $\epsilon = 1.6$ mHa to obtain its dissociation curve.
We estimate each $E(D)$ using \ac{EBS} with parameters $\epsilon = 1.6$~mHa, $\delta = 0.1$ and $\beta = 1.1$ as detailed in \cref{subsec:Bernstein_stopping_rule}.
This procedure is repeated $100$ times independently to gather statistics.
\begin{figure}[b]
\centering
\input{plots/single_shot}
\end{figure}
\end{document}